\documentclass[useAMS,usenatbib]{mn2e}
\usepackage{epsfig}
\usepackage{deluxetable} 
\usepackage[T1]{fontenc}
\usepackage{aecompl}

\title[MWL observations of PMN J0948$+$0022 in 2011]{Multiwavelength observations of the $\gamma$-ray emitting narrow-line Seyfert 1 PMN J0948$+$0022 in 2011}
\author[F. D'Ammando, J. Larsson, M. Orienti, et al.]{F. D'Ammando$^{1,2}$\thanks{E-mail: dammando@ira.inaf.it},
  J. Larsson$^{3}$, M. Orienti$^{1,4}$, C. M. Raiteri$^{5}$, E. Angelakis$^{6}$,
  \newauthor A. Carrami\~nana$^{7}$, L. Carrasco$^{7}$, A. J. Drake$^{8}$, L. Fuhrmann$^{6}$, M. Giroletti$^{1}$,\newauthor
  T. Hovatta$^{9}$, W. Max-Moerbeck$^{9}$, A. Porras$^{7}$,
  A. C. S. Readhead$^{9}$, E. Recillas$^{7}$, \newauthor J. L. Richards$^{10}$ \\
$^{1}$INAF - Istituto di Radioastronomia, Via Gobetti 101, I-40129 Bologna, Italy\\
$^{2}$Dipartimento di Fisica, Universit\`a degli Studi di Perugia, Via A. Pascoli, I-06123 Perugia, Italy \\
$^{3}$KTH, Department of Physics, and the Oskar Klein Centre, AlbaNova, SE-106 91 Stockholm, Sweden \\
$^{4}$Dipartimento di Astronomia, Universit\`a di Bologna, Via Ranzani 1, I-40127 Bologna, Italy \\
$^{5}$INAF - Osservatorio Astrofisico di Torino, Via Osservatorio 20, I-10025 Pino Torinese (TO), Italy \\
$^{6}$Max-Planck-Institute f\"ur Radioastronomie, Auf dem H\"ugel 69, D-53121 Bonn, Germany \\
$^{7}$Instituto National de Astrof\'{i}sica, \'{O}ptical, y Electr\'{o}nica, Tonantzintla, 72840 Puebla, Mexico \\
$^{8}$California Institute of Technology, 1200 E. California Blvd, CA 91225, USA \\
$^{9}$Cahill Center for Astronomy and Astrophysics, California Institute of Technology 1200 E. California Blvd, Pasadena, CA 91125, USA \\
$^{10}$Department of Physics, Purdue University, 525 Northwestern Avenue, West Lafayette, IN 47907, USA
}
\begin{document}

\date{Accepted. Received; in original form}

\maketitle

\label{firstpage}

\begin{abstract}
We report on radio-to-$\gamma$-ray observations during 2011 May--September of
PMN J0948$+$0022, the first narrow-line Seyfert 1 (NLSy1) galaxy detected in
$\gamma$-rays by {\em Fermi}-LAT. Strong variability was observed in
$\gamma$-rays, with two flaring periods peaking on 2011 June 20 and July 28.
The variability observed in optical and near-infrared seems to have no counterpart in $\gamma$-rays. 
This different behaviour could be related to a bending and inhomogeneous jet or a turbulent extreme multi-cell scenario. The radio spectra showed a variability pattern typical of relativistic jets.

\noindent The XMM spectrum shows that the emission from the jet dominates
above $\sim$2 keV, while a soft X-ray excess is evident in the low-energy part
of the X-ray spectrum. Models where the soft emission is partly produced by blurred reflection or Comptonisation of the thermal disc emission
provide good fits to the data. The X-ray spectral slope is similar to that found in radio-quiet NLSy1, suggesting that a standard accretion disc is present, as expected from the high accretion rate. Except for the soft X-ray excess, unusual in jet-dominated AGNs, PMN J0948$+$0022 shows all characteristics of the blazar class.
\end{abstract}

\begin{keywords}
galaxies: active -- galaxies: nuclei -- galaxies: Seyfert -- galaxies:
individual (PMN J0948$+$0022) -- gamma-rays: general
\end{keywords}

\section{Introduction}

Narrow-line Seyfert 1 (NLSy1) galaxies, first suggested as a distinct class of active galactic nuclei (AGNs)
by \citet{osterbrock85}, are characterized in optical by their narrow permitted
emission lines (full-width at half maximum FWHM $\leq$ 2000 km s$^{-1}$), weak
[OIII]$\lambda$5007 emission line ([OIII]/H$\beta$ $<$ 3), and strong Fe\,\textsc{ii}
emission lines. They also exhibit strong X-ray variability, steep X-ray
spectra, substantial soft X-ray excess and relatively high luminosity
\citep{boller96,zhou06}. These characteristics suggest that NLSy1s have smaller
masses of the central black hole (M$_{\rm BH}$ = 10$^{6}$--10$^{8}$
M$_{\odot}$) and higher $\dot{M}/\dot{M}_{Edd}$ values (up to the Eddington limit or above) than those observed in blazars and radio
galaxies. Only a small percentage ($<$7\%) of NLSy1 are radio-loud
\citep{komossa06} compared to $\sim$15\% of the quasars. In the radio-loud
NLSy1s, flat radio spectra and flux density variability suggest that several of them
could host relativistic jets \citep{zhou03, doi06, yuan08}. 

PMN J0948$+$0022 shows optical properties typical of a NLSy1 (i.e.~ FWHM(H$\beta$) = 1432$\pm$87 km s$^{-1}$, [OIII]/H$\beta$$\sim$0.1,
  a strong Fe\,\textsc{ii} bump), and a radio loudness of $R$ = 355 \citep{yuan08}.
High brightness temperature and a compact structure have
been observed for PMN J0948+0022 \citep{doi06}, in addition to a possible
core-jet structure \citep{giroletti11}. This source was
the first radio-loud NLSy1 to be detected in $\gamma$ rays by the Large Area
Telescope (LAT) on board the {\em Fermi Gamma-ray Space Telescope} \citep{abdo09a}. After
that, other 4 radio-loud NLSy1s were detected with high
significance in $\gamma$ rays \citep{abdo09b, dammando12}, suggesting the
radio-loud NLSy1s as a third class of $\gamma$-ray emitting AGN with
relativistic jets. On the contrary, no radio-quiet
Seyfert galaxies have been detected in $\gamma$ rays \citep{ackermann12}. Three $\gamma$-ray flares were observed from PMN
J0948$+$0022 during 2010--2013, reaching daily peak fluxes of (1-2)$\times$10$^{-6}$ photons cm$^{-2}$ s$^{-1}$
\citep{foschini11,dammando11,dammando13}. This indicates that radio-loud
NLSy1s also can host powerful relativistic jets such as blazars.  These findings pose intriguing questions about the nature 
of these $\gamma$-ray emitting NLSy1s, the onset of production of their relativistic jets, and the
cosmological evolution of radio-loud AGNs.

After the discovery of $\gamma$-ray emission from PMN J0948$+$0022, this
source has been the target of different multifrequency campaigns with the aim of
understanding its nature. The first spectral energy distributions (SEDs)
collected for this object, as well as for the other three $\gamma$-ray NLSy1s
detected in the first year of {\em Fermi} operation, showed clear similarities
with blazars: a double-humped shape with a first peak in the IR/optical band
due to synchrotron emission, a second peak in the MeV/GeV band likely due to inverse
Compton emission, and an accretion disk component in UV. The physical
parameters of these NLSy1s are blazar-like, and jet powers are
in the average range of blazars \citep{abdo09b}. In addition, the comparison of the SED of PMN J0948$+$0022 during the 2010 July flaring
activity with that of 3C 273, a typical flat spectrum radio quasar (FSRQ), showed a more extreme Compton
dominance in the NLSy1. The disagreement between the two SEDs can be due to the
difference in black hole (BH) masses and Doppler factor of the two jets
\citep{foschini11}. The radio-to-$\gamma$-ray light curves of the source
collected over years showed correlated variability, with a delay of a few
months of the radio emission with respect to the $\gamma$-ray, as usually
observed in FSRQs \citep{abdo09c, foschini12}.

In this paper, we discuss the results of the analysis of the multiwavelength
data of PMN J0948$+$0022 collected during 2011 May--September. Part of the
data presented here has been already published in \citet{foschini12}. XMM-{\em Newton} and Catalina Real-time Transient Survey data are presented here for the first time. {\em Fermi}--LAT, {\em Swift}, and MOJAVE data are re-analyzed.
The paper is
organized as follows. In Section 2, we report the LAT data analysis and
results, while in Sections 3 and 4 we present the
X-ray and optical/UV results of the {\em Swift} and XMM-{\em Newton}
observations, respectively. Near-infrared (NIR) and optical data from
ground-based observatories are presented and
discussed in Section 5. Radio data
collected by the Very Long Baseline Array (VLBA) interferometer, the
40 m Owens Valley Radio Observatory (OVRO), the 32 m Medicina, the 13.7 m
Mets\"ahovi, and 100 m Effelsberg
single-dish telescopes are presented and discussed in Section 6. In Section 7,
we discuss the properties of the source and draw our conclusions. 

Throughout the paper the quoted uncertainties are
given at the 1$\sigma$ level, unless otherwise stated, and the photon indices are
parameterized as $dN/dE \propto E^{-\Gamma}$, where $\Gamma$ is the photon index. We adopt a $\Lambda$ cold dark matter ($\Lambda$--CDM) cosmology with $H_0$ = 71 km
s$^{-1}$ Mpc$^{-1}$, $\Omega_{\Lambda} = 0.73$, and $\Omega_{\rm m} =
0.27$. The corresponding luminosity distance at $z =0.5846$ is d$_L =  3413$\ Mpc, and 1 arcsec corresponds to a projected size of 6.590 kpc.

\section{{\em Fermi}-LAT data: analysis and results}
\label{FermiData}

\subsection{Observations and data reduction}

The {\em Fermi}-LAT  is a pair-conversion telescope operating from 20 MeV to
$>$ 300 GeV. It has a large peak effective area ($\sim$8000 cm$^{2}$ for 1 GeV
photons), an energy resolution of typically $\sim$10\%, and a field of view of
about 2.4 sr with an angular resolution (68\% containment radius) better than
1$^{\circ}$ for energies above 1 GeV. Further details about the {\em Fermi}-LAT are given in \citet{atwood09}. 

\begin{figure}
\centering
\includegraphics[width=8.0cm]{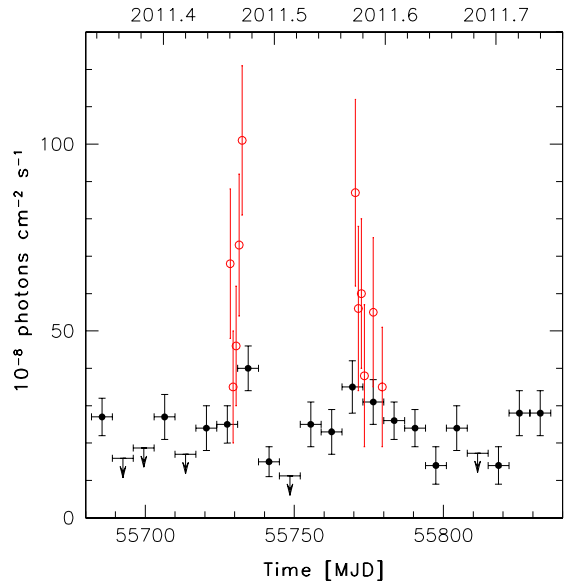}
\caption{Integrated flux light curve of PMN J0948$+$0022 obtained by {\em
    Fermi}-LAT in the 0.1--100 GeV energy range during 2011 May 1 -- September 30 (MJD 55682--55835) with 7-day time
  bins. Arrows refer to 2$\sigma$ upper limits on the source flux. Upper
  limits are computed when $TS$ $<$ 10. Open circles represent daily fluxes
  reported for the periods of high activity.}
\label{Fig1}
\end{figure}

The LAT data reported in this paper were collected from 2011 May 1 (MJD 55682)
to September 30 (MJD 55834). During this time, the {\em Fermi} spacecraft
operated almost entirely in survey mode. The analysis was performed with the
{\small ScienceTools} software package version v9r27p1. The LAT data
were extracted within a region of $20^{\circ}$ radius centred at the
location of PMN J0948$+$0022. Only events belonging to the `Source' class were
used. The time intervals when the rocking angle of the LAT was greater than
52$^{\circ}$ were rejected. In addition, a cut on the
zenith angle ($< 100^{\circ}$) was applied to reduce contamination from
the Earth limb $\gamma$ rays, which are produced by cosmic rays interacting with the upper atmosphere. 
The spectral analysis was performed with the instrument response functions
\texttt{P7SOURCE\_V6} using an unbinned maximum-likelihood method implemented
in the Science tool {\small gtlike}. A Galactic diffuse emission model and isotropic component, which is the sum of
an extragalactic and residual cosmic-ray background, were used to model the background\footnote{http://fermi.gsfc.nasa.gov/ssc/data/access/lat/Background\\Models.html}. The normalizations of both components in the background model were allowed to vary freely during the spectral fitting. 

We analysed a region of interest of $10^{\circ}$ radius centred at the
location of PMN J0948$+$0022. We evaluated the significance of the $\gamma$-ray signal from the sources by
means of a maximum-likelihood test statistic $TS$ = 2 ($logL_1$ -- $logL_0$), where $L$ is the likelihood of the data given the model with ($L_1$) or
without ($L_0$) a point source at the position of PMN J0948$+$0022 \citep{mattox96}. Following the second {\em Fermi} LAT source
catalogue \citep[2FGL;][]{nolan12}, the spectral model used for PMN
J0948$+$0022 is a log-parabola, $dN/dE \propto$ $(E/E_{0})^{-\alpha-\beta \, \log(E/E_0)}$
\citep[]{landau86, massaro04}, where the parameter $\alpha$ is the spectral
slope at the energy $E_0$ and the parameter $\beta$ measures the curvature
around the peak. We fixed the reference energy $E_{0}$ to 272 MeV as in the 2FGL catalogue.
The source model used in {\small gtlike} includes all of the point sources from the 2FGL catalogue that
fall within $15^{\circ}$ of PMN J0948$+$0022. The spectra of these sources
were parametrized by power law functions, $dN/dE \propto$ $(E/E_{0})^{-\Gamma}$,
except for 2FGL J0909.1$+$0121, for which we used a log-parabola as in the 2FGL catalogue. 
A first maximum-likelihood analysis was performed to remove from the model the sources having
$TS$ $<$ 25 and/or a predicted number of counts based on the fitted model
$N_{pred} < 3 $. A second maximum-likelihood analsyis was performed
on the updated source model. The fitting procedure has been performed with the
sources within 10$^{\circ}$ of PMN J0948$+$0022 included with the
normalization factors and the photon indices left as free parameters. For the
sources located between 10$^{\circ}$ and 15$^{\circ}$ from our target, we kept the
normalizations and the photon indices fixed to the values of the 2FGL catalogue.

\subsection{Results}

Integrating over the period 2011 May 1--September 30 (MJD 55682--55835) the fit
yielded a $TS$ = 674 in the 0.1--100 GeV energy range, with an average integral flux of (21.6 $\pm$ 1.5) $\times$10$^{-8}$ ph cm$^{-2}$ s$^{-1}$,
a spectral slope $\alpha$ = 2.41 $\pm$ 0.10, and a curvature parameter around the peak $\beta$ = 0.20 $\pm$
0.06. The corresponding apparent isotropic $\gamma$-ray luminosity is
$\sim$1.8$\times$10$^{47}$ erg s$^{-1}$. As a comparison during the first two
years of {\em Fermi} operation (2008 August 4--2010 August 4) the average
integral flux was (9.2 $\pm$ 0.5) $\times$10$^{-8}$ ph cm$^{-2}$ s$^{-1}$, with a
spectral slope $\alpha$ = 2.26 $\pm$ 0.08, and a curvature parameter around
the peak $\beta$ = 0.26 $\pm$ 0.06 \citep{nolan12}. Thus the average 0.1--100
GeV flux over 2011 May--September is about a factor of 2 higher than the 2FGL
flux, but no significant spectral changes were observed. 

Fig.~\ref{Fig1} shows the $\gamma$-ray light curve for the period considered,
using a log-parabola spectral model and 7-day time bins. For each time bin, the spectral
parameters of PMN J0948$+$0022 and all sources within 10$^{\circ}$ of it were
frozen to have the parameter values resulting
from the likelihood analysis over the entire period. For the highest
significance periods, we also reported as open circles the fluxes in 1-day time intervals.
If $TS$ $<$ 10, 2$\sigma$ upper limits were calculated. All uncertainties in
measured $\gamma$-ray flux and index reported throughout the paper are statistical only.
The systematic
uncertainty  in the effective area is energy dependent: it amounts to $10\%$ at 100 MeV, decreasing to
$5\%$ at 560 MeV, and increasing to $10\%$ above 10 GeV
\citep{ackermann12}.

A daily peak flux of (101 $\pm$ 20) $\times$10$^{-8}$ ph cm$^{-2}$ s$^{-1}$ in the 0.1--100 GeV energy range was detected on 2011
June 20 (MJD 55732), representing an increase of a factor of 5 with respect to
the 2011 May--September average flux and more than an order of magnitude above the average 2FGL
flux of the source. The corresponding apparent isotropic $\gamma$-ray
luminosity is $\sim$8.8$\times$10$^{47}$ erg s$^{-1}$. Preliminary results
about this $\gamma$-ray flaring activity were reported by \citet{dammando11} and \citet{lucarelli11}. A second
$\gamma$-ray flaring activity was observed at the end of 2011 July,
peaking on July 28 (MJD 55770) with a flux of (87 $\pm$ 25) $\times$10$^{-8}$ ph cm$^{-2}$ s$^{-1}$ in the 0.1--100 GeV energy range.
By means of the {\small gtsrcprob} tool, we estimated that
the highest energy photon emitted from PMN J0948$+$0022 (with probability $>$ 80\% of being associated with the source) was
observed on 2011 September 13 (MJD 55807), at a distance of 0.09$^{\circ}$ from
the source and with an energy of 3.2 GeV. This suggests that this NLSy1 emits
mainly at E $<$ 10 GeV, even during flaring activity \citep[see e.g.,][]{foschini11}.

\begin{table*}
\caption{Log and fitting results of {\em Swift}-XRT observations of PMN
  J0948$+$0022 using an absorbed power law model with an absorbing column
  density of $N_{\rm H}$ = 5.07$\times$10$^{20}$ cm$^{-2}$.}
\begin{center}
\begin{tabular}{cccccc}
\hline \hline
\multicolumn{1}{c}{Date} &
\multicolumn{1}{c}{Date} &
\multicolumn{1}{c}{Net exposure time} &
\multicolumn{1}{c}{Photon index} &
\multicolumn{1}{c}{Flux 0.3--10 keV$^{a}$} \\
\multicolumn{1}{c}{(UT)} &
\multicolumn{1}{c}{(MJD)} &
\multicolumn{1}{c}{(sec)} &
\multicolumn{1}{c}{($\Gamma$)} &
\multicolumn{1}{c}{($\times$10$^{-12}$ erg cm$^{-2}$ s$^{-1}$)} \\
\hline
2011-Apr-29 & 55680 & 1978 & $1.81 \pm 0.19$ & $4.3 \pm 0.4$ \\
2011-May-15 & 55696 & 4657 & $1.75 \pm 0.15$ & $4.5 \pm 0.3$ \\
2011-May-28 & 55709 & 3629 & $1.80 \pm 0.16$ & $4.8 \pm 0.5$ \\
2011-June-04 & 55716 & 2020 & $1.76 \pm 0.20$ & $3.5 \pm 0.4$ \\
2011-June-14 & 54726 & 5160 & $1.65 \pm 0.16$ & $3.7 \pm 0.3$ \\
2011-July-02 & 54744 & 2008 & $1.44 \pm 0.17$ & $5.9 \pm 0.6$ \\
\hline
\hline
\end{tabular}
\end{center}
\label{XRT}
$^{\rm a}$Unabsorbed flux
\end{table*}

\section{{\em Swift} Data: Analysis and Results}
\label{SwiftData}

The {\em Swift} satellite \citep{gehrels04} performed six observations
of PMN J0948$+$0022 between 2011 April 29 and July 2. The observations were
performed with all three on board instruments: the Burst Alert Telescope
\citep[BAT;][15--150 keV]{barthelmy05}, the X-ray Telescope
\citep[XRT;][0.2--10.0 keV]{burrows05}, and the Ultraviolet/Optical Telescope
\citep[UVOT;][170--600 nm]{roming05}. The hard X-ray flux of this source is below the sensitivity
of the BAT instrument for the short exposures of these observations and therefore the data from this instrument were not used.
Moreover, the source is not present in the {\em Swift}-BAT 70-month hard X-ray catalogue \citep{baumgartner13}.

\subsection{{\em Swift}-XRT}

The XRT data were processed with standard procedures ({\small xrtpipeline v0.12.6}), filtering, and screening criteria by using the {\small HEASoft}
package (v6.12). The data were collected in photon counting mode for all of the observations. The source count rate
was low ($<$ 0.5 counts s$^{-1}$); thus pile-up correction was not
required. Source events were extracted from a circular region with a radius of
20 pixels (1 pixel $\sim$ 2.36 arcsec), while background events were extracted from a circular region with radius of 50 pixels away from the source
region or other bright sources. Ancillary response files were generated with \texttt{xrtmkarf} and
account for different extraction regions, vignetting and point spread function
corrections. We used the spectral redistribution matrices in the
Calibration data base maintained by {\small HEASARC}. When the number of photons
collected was too low ($<$ 200 counts), the spectra
were rebinned with a minimum of 1 count per bin and the Cash statistic
\citep{cash79} was used. Since the effective area of {\em Swift}-XRT is a
factor of $\sim$10 lower than that of the XMM-{\em Newton} EPIC cameras,
detailed spectral modeling was not performed with the XRT observations.
The spectrum was fitted with an absorbed power law using the photoelectric
absorption model {\small TBABS} \citep{wilms00}, with a neutral hydrogen column density fixed
to its Galactic value \citep[5.07$\times$10$^{20}$ cm$^{-2}$;][]{kalberla05}.
The fit results are reported in Table~\ref{XRT}. The relatively harder X-ray spectrum observed for PMN J0948$+$0022
with respect to the other NLSy1s \citep[e.g.,][]{grupe10,Zhou2010} could be due to the
contribution of inverse Compton radiation from a relativistic jet, similarly
to what is found in FSRQs. A photon index 1.4--1.5 was observed
in X-rays also for another $\gamma$-ray NLSy1, SBS 0846$+$513 \citep{dammando12,dammando13b}.

\subsection{{\em Swift}-UVOT}

During the {\em Swift} pointings the UVOT instrument observed PMN J0948$+$0022
in its optical ($v$, $b$, and $u$) and UV ($w1$, $m2$, and $w2$) photometric
bands. The data analysis was performed using the \texttt{uvotsource} task
included in the {\small HEASoft} package (v6.12). Source counts were extracted
from a circular region of 5 arcsec radius centred on the source, while
background counts were derived from a circular region of 10 arcsec radius in the  source
neighbourhood. We calculated the effective wavelengths, count rate to flux
conversion factors, and Galactic extinctions for the UVOT bands according to the procedure explained in
\citet{raiteri10,raiteri11} and \citet{dammando12}.
The observed magnitudes are reported in Table~\ref{UVOT}, and flux
densities for the $v$, $u$, and $w2$ filters are shown in Fig.~\ref{MWL}. The optical
$u$-band magnitude ranged from 17.67 to 17.11, with the peak detected on 2011 June 4 (MJD 55716). No
significant variability was observed in UV, but we noted that no UV observations were available
during the optical peak in $u$-band.

\begin{table*}
\caption{Results of the {\em Swift}-UVOT observations of PMN J0948$+$0022.}
\begin{center}
\begin{tabular}{cccccccc}
\hline \hline
\multicolumn{1}{c}{Date (UT)} &
\multicolumn{1}{c}{Date (MJD)} &
\multicolumn{1}{c}{$v$} &
\multicolumn{1}{c}{$b$} &
\multicolumn{1}{c}{$u$} &
\multicolumn{1}{c}{$w1$} &
\multicolumn{1}{c}{$m2$} &
\multicolumn{1}{c}{$w2$} \\
\hline
2011-Apr-29 & 55680 &  -- &  --           & 17.27$\pm$0.03 & -- & -- & -- \\
2011-May-15 & 55696 &  18.20$\pm$0.13 & 18.43$\pm$0.08 & 17.67$\pm$0.07 & 17.46$\pm$0.05 & 17.44$\pm$0.06 & 17.46$\pm$0.04 \\
2011-May-28 & 55709 &  18.05$\pm$0.15 & 18.31$\pm$0.09 & 17.48$\pm$0.07 & 17.31$\pm$0.05 & 17.35$\pm$0.06 & 17.33$\pm$0.04 \\
2011-June-04 & 55716 &  -- &  --           & 17.11$\pm$0.03 & -- & -- & -- \\
2011-June-14 & 55726 &  17.86$\pm$0.11 &  18.31$\pm$0.08 & 17.34$\pm$0.06 & 17.34$\pm$0.05 & 17.35$\pm$0.05 & 17.29$\pm$0.04 \\
2011-July-02 & 55744 &  -- & --            & 17.64$\pm$0.30 & -- & -- & -- \\
\hline
\hline
\end{tabular}
\end{center}
\label{UVOT}
\end{table*}

\section{XMM-{\em Newton} data: analysis and results}
\subsection{Observations and data reduction}
\label{observations}

XMM-{\em Newton} \citep{jansen01} observed PMN~J0948+0022 on 2011 May
28-29 (MJD 55709--55710) for a total duration of 93 ks (observation ID
067370101, PI: D'Ammando). A simultaneous observation was performed by {\em Swift} on 2011 May 28.

The EPIC pn was operated in the large window mode and the EPIC MOS cameras (MOS1 and MOS2) were operated in
the prime partial mode. The data were reduced using the XMM-{\em Newton}
Science Analysis System ({\small SAS v11.0.0}), applying standard event
selection and filtering. Inspection of the background light curves showed that strong flaring was present during the whole observation.  The flaring time intervals were removed for the spectral analysis, leaving good exposure times of 36, 58 and 60 ks for the pn, MOS1 and MOS2, respectively. For each of the detectors the source spectrum was extracted from a circular region of radius 34 arcsec centred on the source, and the background spectrum from a nearby region on the same chip. All the spectra were binned to contain at least 20 counts per bin to allow for $\chi^2$ spectral fitting. 

The Optical Monitor (OM) and Reflection Grating Spectrometers (RGS1 and RGS2)
were also operating during the observation. No spectral lines were detected in
the RGS spectra. This could be because the signal was rather
low. The RGS data are not discussed further in this work. The data from the OM are discussed
in Sect.~\ref{OM}.

\subsection{X-ray spectral analysis}\label{XMM}

All spectral fits were performed over the 0.3-10~keV energy range using
{\small XSPEC v.12.7.1}. The energies of spectral features are quoted for the
rest frame of the source while plots are in the observed frame, unless
otherwise stated. All errors are quoted for 90\% confidence for one
interesting parameter ($\Delta \chi^2 = 2.7$). The data from the three EPIC cameras were
initially fitted separately, but since good agreement was found  ($<5\%$) we
proceeded to fit them together.  As for the {\it Swift} spectra, Galactic
absorption was included in all fits using the {\small TBABS} model. The results of all the fits are presented in Table \ref{xmmfits}. It is clear
that a simple power law model was insufficient to describe the data, while a
broken power law yielded an acceptable fit (the improvement between the models
is $\Delta \chi^2 = 969$ for two additional free parameters). The broken
power law has a break at observed energy $\rm{E_{break}}=1.72^{+0.09}_{-0.11}$
~keV, with photon indices below and above the break of $\Gamma_1 =
2.14^{+0.03}_{-0.02}$ and $\Gamma_2 = 1.48^{+0.04}_{-0.03}$, respectively. In
order to check for intrinsic absorption, a neutral absorber at the redshift of
the source was added to this model, but it was found not to be required. The
power law slope above the break energy is significantly harder than those observed
in radio-quiet NLSy1s \citep[e.g.,][]{grupe10,Zhou2010}, but similar to the slopes found
in FSRQs \citep[e.g.,][]{Donato2001}. As already noted based on the {\em Swift}-XRT observations above, this makes highly probable that the emission
from the jet dominates the X-ray spectrum at these energies. This conclusion is further supported by the 
fact that there was no detection of an Fe line in the spectrum. The 90\% upper limit on the equivalent width (EW) of a narrow emission line at 6.4~keV is EW$<19$~eV, or EW$<29$~eV if the energy is allowed to vary between 6.4 and 7 keV. 

\begin{figure}
\begin{center}
\rotatebox{270}{\resizebox{!}{80mm}{\includegraphics{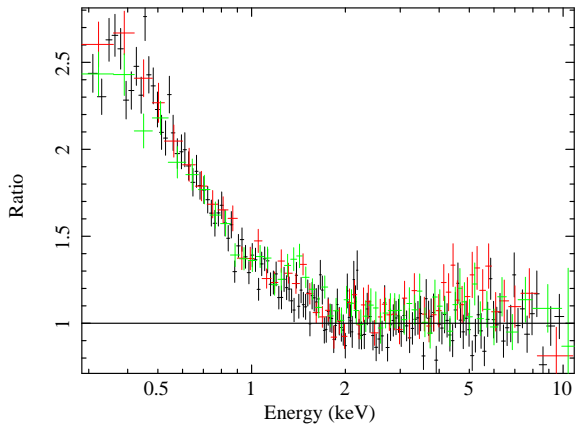}}}
\caption{\small{XMM-{\em Newton} EPIC pn (black), MOS1 (red) and MOS2 (green) data shown as a ratio to  a power law with $\Gamma = 1.48$.}}
\label{softexcess}
\end{center}
\end{figure}

On the other hand, the steeper slope found at low energies may be associated
with the corona and the accretion disc of the system. In radio-quiet NLSy1s, the
low-energy part of the spectrum is composed of a steep power law, originating
in the corona, as well as a strong so-called soft excess, the origin of which
is debated. Fig.~\ref{softexcess} shows the spectrum of PMN J0948$+$0022 as a ratio to the power law with $\Gamma_2 = 1.48^{+0.04}_{-0.03}$, which fits the spectrum above the break energy, illustrating the strength of the soft component in this source.

 \begin{figure*}
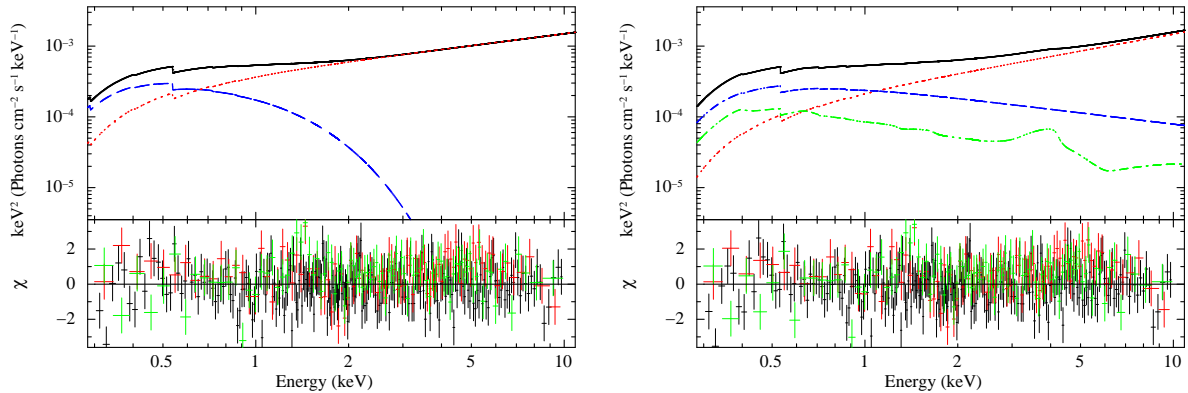

\begin{center}
\rotatebox{270}{\resizebox{!}{80mm}{\includegraphics{comp_model.eps}}}
\rotatebox{270}{\resizebox{!}{80mm}{\includegraphics{ref_model.eps}}}
\caption{\small{Components of the best-fitting models from Table~\ref{xmmfits} together with the residuals of the fits.  Left: Comptonised blackbody (blue, dashed line), power law (red, dotted line) and their sum (black, solid line). Right: Steep power law (blue, dashed line), its associated reflection spectrum (green, dashed-dotted line), hard power law (red, dotted line) and their sum (black, solid line). In the lower panels the black, red, and green points represent data from pn, MOS1 and MOS 2, respectively.}}
\label{xmm_models}
\end{center}
\end{figure*}
 
The soft component can be fitted with a blackbody model with kT $\sim0.18$~keV,
which is within the typical range of temperatures for the soft excess
\citep{Gierlinski2004}. However, it is well known that such a high temperature
is inconsistent with standard accretion disc theory, which predicts that the
thermal emission from the disc around a super-massive BH should peak in
the ultraviolet band. Indeed, from the SED of the
source \citep[see e.g.,][]{foschini11,foschini12} the observed emission from the accretion
disc seems to peak at 20~eV, corresponding to a disc temperature of 11~eV in the rest frame of the source.
  
\noindent The soft X-ray emission may instead be due to Comptonisation of the disc
emission by a population of electrons with low temperature and large optical
depth, which may exist in a transition region between the disc and the corona
\citep{Done2012}. To investigate this model we proceeded to fit the spectrum with a power law
and a Comptonised blackbody, using the {\small COMPTT} model \citep{Titarchuk1994} and
fixing the seed temperature at 11~eV. As seen in Table~\ref{xmmfits} and the
left panel of Fig.~\ref{xmm_models}, this model provides a good fit to the
spectrum, with best-fit electron temperature $\rm{kT_e} =
0.50^{+0.16}_{-0.09}$~keV and optical depth $\tau=10.2^{+0.3}_{-0.1}$.  We
note that allowing a higher seed temperature of 20~eV changes only the
normalisation of the Comptonised component, without significantly affecting
any of the other best fit values presented in Table \ref{xmmfits}. The photon
index of the power law in this model is $\Gamma_2$ = 1.44$\pm$0.03. In
principle the model should also contain a second power law to account for any
emission from the corona, but such a component was not required in
the fits.

 An alternative explanation for the soft excess is that it is due to
 relativistically blurred reflection from the accretion disc. Such a model can
 explain the fact that the energy of the soft excess is observed to be fairly
 constant for black holes spanning several orders of magnitude in mass
 \citep{Gierlinski2004,Bianchi2009}, and has been successfully applied for a
 large number of sources \citep{Crummy2006}. In this picture the X-ray
 spectrum of PMN~J0948$+$0022 would be composed of  a steep power law
 associated with the corona, a reflection component resulting from irradiation
 of the disc by this power law, as well as a hard power law associated with
 the jet, dominating at high energies. The large number of parameters in such
 a model are impossible to constrain with the available data, especially given
 that there is no Fe line or Compton hump for the reflection model to anchor
 to. To set some constraints on such a scenario we constructed a model made of two power laws (with the photon indices constrained to be steeper
 and flatter than 2, for the corona and jet, respectively) together with a
 relativistically blurred reflection component with parameters fixed at `standard' values. The reflection was modelled with {\small KDBLUR} acting
 on the {\small REFLIONX} model by \cite{Ross2005}, with the photon index of
 the illuminating power law tied to the steep power law. The inner and outer
 radii of the disc were fixed at 6 gravitational radii ($r_{\rm{g}}$) and 400~$r_{\rm{g}}$,
 respectively. Solar abundances were assumed and the emissivity index was
 fixed at $q=3$ (i.e. assuming a standard emissivity profile of $r^{-3}$ from
 a central point source). The
 inclination was fixed at $i = 3^{\circ}$, which is the angle of the jet inferred
 by \cite{foschini11}. The disc may not be exactly perpendicular to the jet,
 but we note that allowing a larger value, i.e. up to around $i = 15^{\circ}$, did not change the results. After optimisation the ionisation state of the disc 
was fixed at $\xi  = 3000\  \rm{erg\ cm\ s^{-1}}$. As seen in
Table~\ref{xmmfits} this model results in a fit of quality comparable to the Comptonised blackbody. However, it should be noted that it is
the steep power law and not the reflection component that provides most of the
flux at low energies, as seen in Fig.~\ref{xmm_models}, which shows the model
components. In fact, a fit with two power laws but no reflection is only
marginally worse ($\Delta\chi^2$ = 6 for one more degree of freedom, see Table
\ref{xmmfits}). Allowing each of the parameters of the reflection model to
vary did not significantly change these results.

\begin{figure}
\begin{center}
\rotatebox{270}{\resizebox{!}{80mm}{\includegraphics{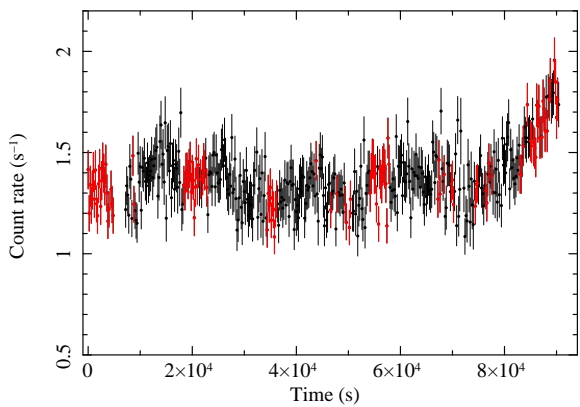}}}
\caption{\small{0.2--10~keV EPIC pn light curve on a 200 s time scale, using only fully exposed bins. The time periods used for the spectral analysis are marked in red. The remaining time intervals were severely affected by background flaring.}}
\label{xmmlc}
\end{center}
\end{figure}

 \begin{table}
\begin{center}
\begin{tabular}{lll}
\hline\hline \\ [-4pt]
Model& Parameter & Value
 \\[4pt]  
\hline\\[-6pt]

power law & $\Gamma$                                 & $1.88\pm 0.01$  \\
                    & Norm                 &  $6.53\pm 0.05 \times 10^{-4}$     \\
                    &  $\chi^2/\rm{d.o.f.}$          &    2342/1254  \\
\hline 
Broken power law & $\Gamma_1$               & $2.14^{+0.03}_{-0.02}$   \\
& $\rm{E_{break}}$ (keV)                     & $1.72^{+0.09}_{-0.11}$   \\
& $\Gamma_2$                                & $1.48^{+0.04}_{-0.03}$  \\
& Norm                                         & $6.28\pm 0.06 \times 10^{-4}$   \\
& $\chi^2/\rm{d.o.f.}$                      &  1373/1252\\
\hline 
power law  +               & $\Gamma$              		   & $1.44\pm0.03$   \\
           & PL Norm          & $4.13\pm 0.17 \times 10^{-4}$   \\
 compTT		           & $\rm{T_0}$ (keV)     		   & $1.10\times 10^{-2 f}$   \\
                           & $\rm{kT_e}$ (keV)     		   & $0.50^{+0.16}_{-0.09}$\\
			   & $\tau$ 			           & $10.2^{+0.3}_{-0.1}$\\
 		           & Comp Norm			           & $0.38^{+0.07}_{-0.06}$ \\	
		           & $\chi^2/\rm{d.o.f.}$                  &  1329/1251\\
\hline
power law (jet) + &       $\Gamma_{\rm{j}}$       		   &    $1.21^{+0.07}_{-0.04}$ \\
 	          & PL $\rm{Norm _{j}}$                               & $2.4^{+0.4}_{-0.2} \times 10^{-4}$   \\
power law (corona) + &   $\Gamma_{\rm{c}}$       		  &  $2.53^{+0.14}_{-0.09}$  \\
 	             & PL $\rm{Norm _{c}}$                            & $2.7^{+0.7}_{-0.4} \times 10^{-4}$   \\
Blurred reflection  &  q                                          & $3^{f}$\\
			        &  $R_{\rm{in}}\  (r_{\rm{g}})$                                         & $6^{f}$ \\
			         &  $R_{\rm{out}}\ (r_{\rm{g}})$                                      & $400^{f}$\\
			         &  incl    (deg)                                            & $3^{f}$    \\
			          &  Fe/solar                                               &  $1^{f}$ \\
			        &  $\xi \  \rm{(erg\ cm\ s^{-1})}$             & $3000^{f}$ \\
			        & Ref Norm                                                    & $2.3^{+1.2}_{-0.9} \times 10^{-9}$     \\

   &  $\chi^2/\rm{d.o.f.}$                   &    1333/1251  \\
   \hline
power law (jet) + &       $\Gamma_{\rm{j}}$       		     &    $1.20^{+0.08}_{-0.09}$ \\
 			        & PL $\rm{Norm_{j}}$      			     & $2.4\pm 0.4 \times 10^{-4}$   \\
power law (corona) &   $\Gamma_{\rm{c}}$       		  &  $2.63^{+0.11}_{-0.10}$  \\
 			        & PL $\rm{Norm_{c}}$                                   & $3.7\pm 0.05 \times 10^{-4}$   \\
			        &  $\chi^2/\rm{d.o.f.}$                   &    1339/1252  \\
\hline\hline
\end{tabular}
\caption{\label{xmmfits}\small{Summary of fits to the 0.3--10~keV XMM-{\em Newton} spectra.  All
    fits also included absorption fixed at the Galactic value. Superscript $f$ indicates that a parameter was kept fixed. See text for a description of the models.}}
\end{center}
\end{table}

 \subsection{X-ray variability}
 
\begin{figure}
\begin{center}
\rotatebox{270}{\resizebox{!}{80mm}{\includegraphics{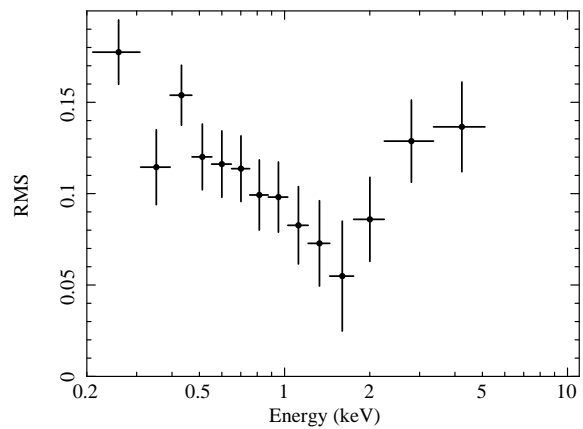}}}
\caption{\small{RMS variability spectrum of the XMM-{\em Newton} observation of PMN~J0948$+$0022, calculated using 0.5~ks bins. }}
\label{rms}
\end{center}
\end{figure}

The EPIC pn light curve is shown in Fig.~\ref{xmmlc}, with the time intervals
with low background that were used for the spectral analysis marked in red. The source was only mildly
variable during the XMM observation, as suggested by the count rate light curve. The observed fluxes corrected for Galactic
absorption  are $F_{2-10~\rm{keV}} = (2.56 \pm 0.05) \times
10^{-12}\  \rm{erg\ cm^{-2}\ s^{-1}}$ and $F_{0.3-10~\rm{keV}} =
4.59^{+0.03}_{-0.05} \times 10^{-12}\  \rm{erg\ cm^{-2}\ s^{-1}}$, in
agreement with the {\em Swift}-XRT results (see Table~\ref{XRT}). 

The root-mean-square (RMS) variability spectrum of the observation is shown in
Fig.~\ref{rms}, and it is calculated following \citet{Vaughan2003} using light
curves with $0.5~$ks bins. The RMS spectrum shows the variability amplitude of
the source as a function of energy, corrected for the variance due to
measurement errors and normalised by the mean count rate in each energy band.
The error bars represent the uncertainty expected from the Poisson noise
\citep{Vaughan2003}. Since the flaring background becomes increasingly
dominant with energy we plot the RMS spectrum only up to $5$~keV (at which
energy the ratio of signal to background is about 4). The variability clearly decreases with energy up to around $1.7~$keV, but then
starts to increase again. It is interesting to note that this break coincides with the break energy of
the broken power law model (see Table~\ref{xmmfits}), since it is
  consistent with our interpretation that the Seyfert and jet components
  dominate at low and high energies, respectively (see further Section \ref{xspectrum}).
The RMS spectrum presented in \cite{Bhattacharyya2013} has a different energy
binning, but exhibits an overall shape that is consistent with our
results. While we find that the increase in RMS variability above $\sim 2$~keV is not affected by our choice of
energy binning, it will be important to confirm these results with an
observation where the variability analysis can be extended to higher energies.

\begin{figure}
\begin{center}
\includegraphics[width=8.0cm]{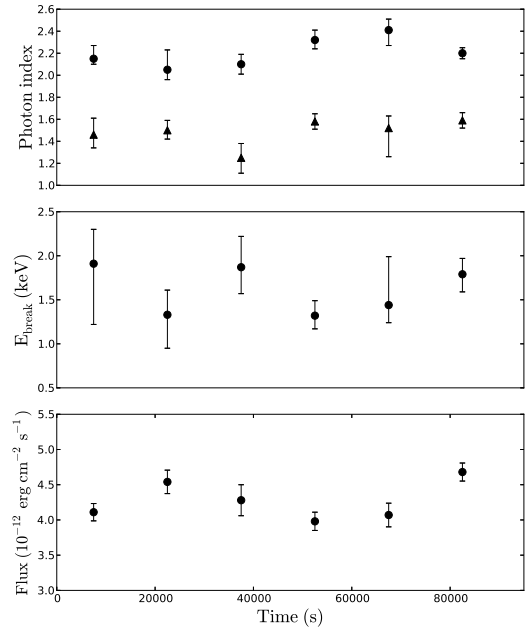}
\caption{\small{Results of fits to time-resolved spectra of the XMM-{\em
      Newton} observation using a broken power law model. The top panel shows
    the temporal evolution of the photon indices (the indices below and above the break energy are shown as filled
circles and filled triangles, respectively), the middle panel shows the break
energy, and the bottom panel shows the $0.3-10$~keV unabsorbed flux derived from the fits.}}
\label{timeresfits}
\end{center}
\end{figure}

We further investigated the spectral variability by fitting the broken power law model to spectra extracted in 15~ks intervals (corresponding to
3--7.5~ks of good exposure time per spectrum). Fig.~\ref{timeresfits}
  shows the evolution of the best-fit parameters and the flux as a function of
  time. There are no systematic trends with time and only the flux is found to
  vary significantly above the 3$\sigma$ level.  However, the photon index and
  $E_{\rm{break}}$ are found to be marginally variable ($>$ $90\%$ CL).

\subsection{Optical Monitor data}
\label{OM}

The Optical Monitor \citep[OM;][]{mason01} on board XMM-{\em Newton} is a 30 cm
telescope carrying six optical/UV filters, and two grisms. Observations of PMN J0948$+$0022 in 2011 May 28--29 consisted of seven subsequent exposures in $v$-band,
followed by ten in $b$-band, nine in $u$-band, ten in $w1$-band and $m2$-band, then five in $w2$-band. All optical exposures were 800 s long, while UV
exposures were 1600 s and 2700 s long. We used the SAS task \texttt{omichain} to
reduce the data and the tasks \texttt{omsource} and \texttt{omphotom} to
derive the source magnitude. Average observed magnitudes are: $v$ = 18.28$\pm$0.06, $b$
= 18.63$\pm$0.02, $u$ = 17.64$\pm$0.02, $w1$ = 17.27$\pm$0.02, $m2$ =
17.24$\pm$0.03, and $w2$ = 17.22$\pm$0.11. The difference of 0.2--0.3 mag in
the optical filters with respect to the {\em Swift}-UVOT observations performed on 2011
May 28 at least partially could be due to the source variability.

\section{Ground-based optical and infrared observations}

\subsection{CRTS}

The source has been monitored by the Catalina Real-time Transient Survey (CRTS)\footnote{http://crts.caltech.edu} \citep{drake09, djorgovski11},
using the 0.68 m Schmidt telescope at Catalina Station, AZ, and an unfiltered
CCD. The typical cadence is four exposures separated by 10 min in a
given night; this may be repeated up to four times per lunation, over a period of
$\sim$6--7 months each year, while the field is observable.  Photometry is obtained using the standard Source-Extractor
package \citep{bertin96}, and transformed from the unfiltered instrumental
magnitude to Cousins $V$ by $V$ = $V_{\rm CSS}$ + 0.31($B-V$)$^{2}$ + 0.04 with
a scatter of 0.056
mag \footnote{http://nesssi.cacr.caltech.edu/DataRelease/FAQ2.html\#improve}. The flux densities
collected by CRTS in $V$ band are reported in Fig.~\ref{MWL}.

\subsection{INAOE}

NIR observations of PMN J0948$+$0022 were performed during 2011 April--May, at
the 2.1 m
telescope ``Guillermo Haro'', with the NIR camera ``CANICA'' equipped with a
Rockwell 1024 $\times$ 1024 pixel Hawaii infrared array, working at 75.4 K, with
standard $J$(1.164-1.328 $\mu$m), $H$(1.485-1.781 $\mu$m), and $Ks$ (1.944-2.294 $\mu$m)
filters. The plate scale is 0.32 arcsec pixel$^{-1}$. Observations were carried out
in series of 10 dithered frames in each filter. 
Data sets were co-added after correcting for bias and flat fielding. Flats
were obtained from sky frames derived from the dithered ones. Magnitudes in $J$,
$H$, and $Ks$ filters are reported in Table~\ref{inaoe}. The flux densities
collected in $J$ and $H$ band are reported also in Fig.~\ref{MWL}.

\begin{table}
\begin{center}
\caption{Results of the INAOE observations of PMN J0948$+$0022 in $J$, $H$, and $K$ bands. \label{inaoe}}
\begin{tabular}{cccc}
\hline \hline
Date & J & H & Ks \\
(MJD)  & (mJy) & (mJy) & (mJy) \\
\hline
55677.219 & $0.864 \pm 0.026$ & $1.248 \pm 0.075$ & $1.234 \pm 0.136$ \\
55689.243 & -- & $0.721 \pm 0.065$ & -- \\
55693.171 & -- & $0.400 \pm 0.020$ & -- \\
55694.139 & $0.469 \pm 0.028$ & $0.601 \pm 0.024$ & $0.929 \pm 0.084$\\
55696.181 & -- & $0.380 \pm 0.030$ & \\
55702.163 & $0.316 \pm 0.016$ & $0.331 \pm 0.020$ & $0.762 \pm 0.084$\\
55703.139 & $0.866 \pm 0.069$ & $1.530 \pm 0.138$ & \\
\hline \hline
\end{tabular}
\end{center}
\end{table}

\section{Radio observations}

\subsection{Effelsberg 100 m}

The radio spectrum of PMN J0948$+$0022 was observed with the Effelsberg
100 m telescope between 2011 May 24 and October 1 within the framework of a {\em Fermi}-related monitoring programme of $\gamma$-ray blazars \citep[F-GAMMA programme;][]{fuhrmann07}. The measurements were conducted with the secondary
focus heterodyne receivers at 2.64, 4.85, 8.35, 10.45, 14.60, 23.05, and 32.00 GHz. The
observations were performed quasi-simultaneously with cross-scans, that is,
slewing over the source position, in azimuth and elevation directions, with
adaptive numbers of sub-scans for reaching the desired sensitivity \citep[for
  details, see][]{fuhrmann08, angelakis08}. Corrections for pointing offset,
gain, atmospheric opacity, and sensitivity have
been applied to the data. The different spectra collected by Effelsberg are represented
in Fig.~\ref{Eff}. 

\begin{figure*}
\centering
\includegraphics[width=15cm]{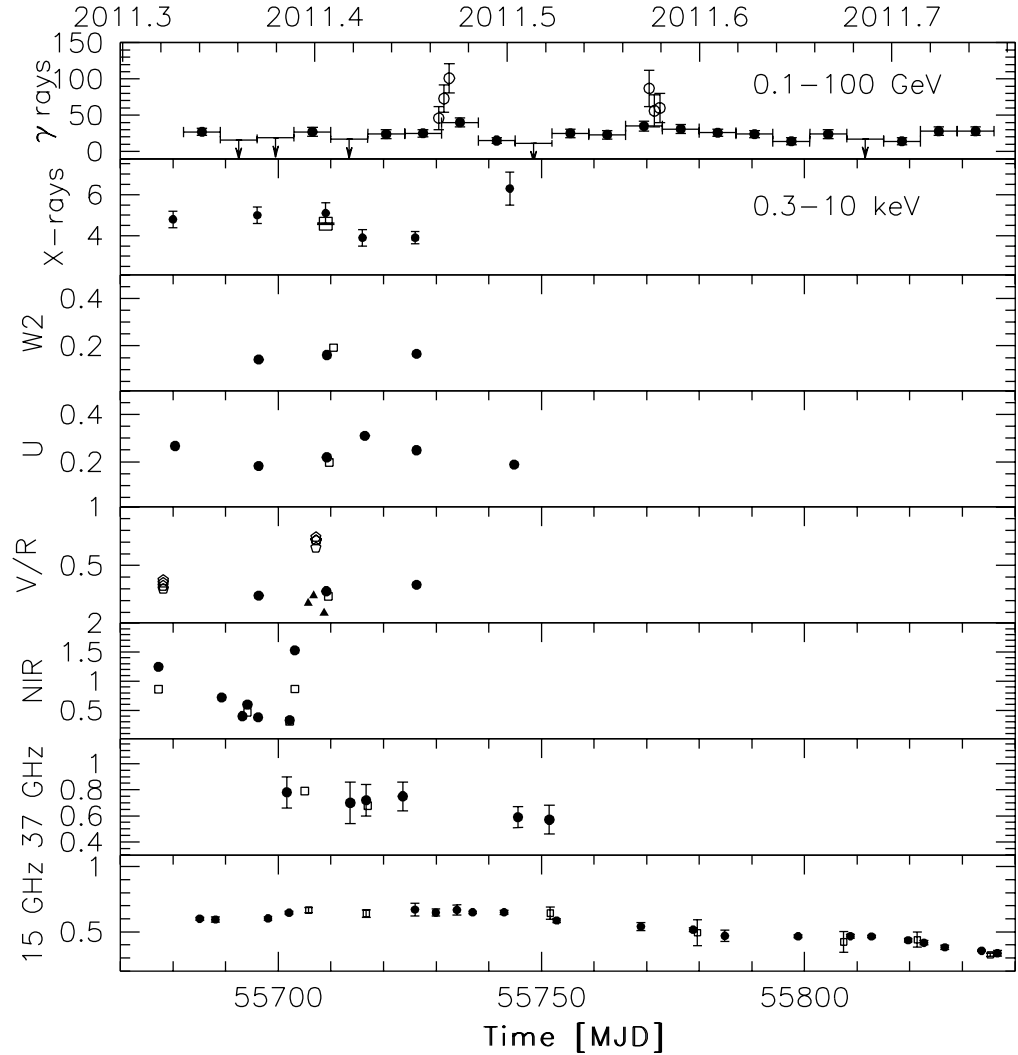}
\caption{Multifrequency light curve of PMN J0948$+$0022 for the period 2011 May 1--September 30 (MJD 55682--55834) collected (from
  top to bottom) in: $\gamma$ rays by {\em Fermi}-LAT (0.1--100 GeV; in units
  of 10$^{-8}$ ph cm$^{-2}$ s$^{-1}$); X-rays by {\em Swift}-XRT (filled circles) and XMM-{\em Newton} (open square) (0.3--10 keV; in units of
  10$^{-12}$ erg cm$^{-2}$ s$^{-1}$); $w2$ band by {\em Swift}-UVOT (filled
  circles) and XMM-OM (open square); $u$ band by {\em Swift}-UVOT (filled
  circles) and XMM-OM (open square); $V$ band by {\em Swift}-UVOT (filled circles), XMM-OM (open square) and CRTS
  (open pentagons), and $R$ band taken from \citet{eggen13} (filled triangles); $J$
  (open squares) and $H$ (filled circles) bands by INAOE; 37
  GHz by Mets\"ahovi (filled circles) and 32 GHz by Effelsberg (open squares); 15 GHz by OVRO (filled circles) and Effelsberg (open
  squares). The flux densities collected from $w2$ to 15 GHz are reported in units of mJy. In the top panel daily integrated $\gamma$-ray fluxes are reported as open circles.}
\label{MWL}
\end{figure*}

\begin{figure}
\centering
\includegraphics[width=7.5cm]{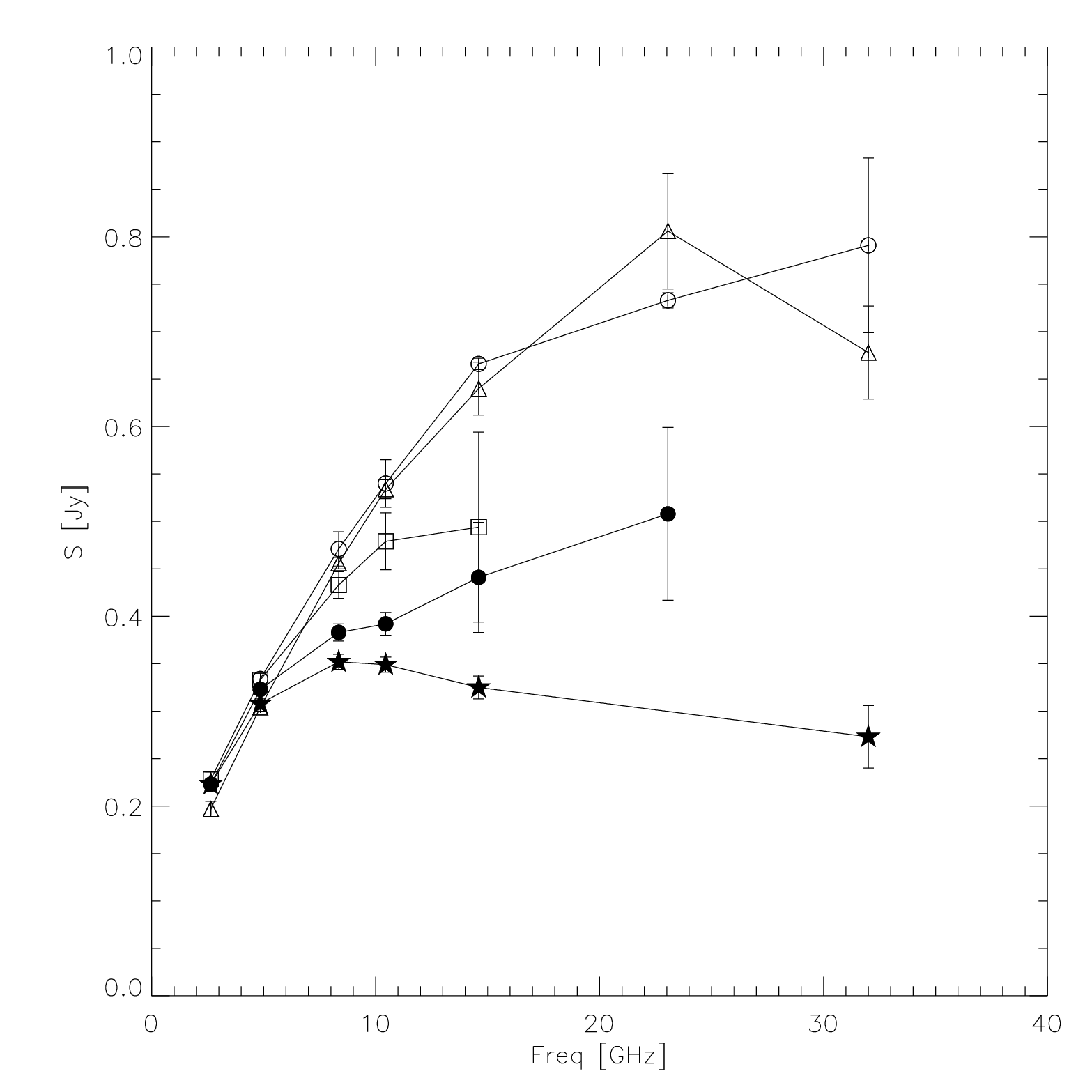}
\caption{Radio spectra of PMN J0948$+$0022 collected from 2.64 and 32 GHz by
  Effelsberg in 5 epochs: 2011 May 24 (MJD 55705; empty circles), 2011 June 5
  (MJD 55717; empty triangles), 2011 August 6 (MJD 55779; empty squares), 2011 September 17 (MJD 55821; filled circles),
  and 2011 October 1 (MJD 55835; filled stars).}
\label{Eff}
\end{figure}

\noindent Radio spectra and fluxes indicate that the source was highly active
already in 2011 May 24 (MJD 55705), before the first peak of the $\gamma$-ray activity,
followed by a flux decrease and a flattening of the spectrum in 2011
August--October. The spectral index calculated between 8.4 and 32 GHz changes between
$-$0.4$\pm$0.1 and 0.2$\pm$0.1 from 2011 May to October. No
significant flux changes were observed at frequencies below 8.4 GHz. This is likely due
to opacity effects at the low frequencies. Flux densities at 32 GHz and 14.6 GHz
are also shown in Fig.~\ref{MWL}.

\subsection{Mets\"ahovi}

Observations at 37 GHz were made with the 13.7 m Mets\"ahovi radio
telescope, which is a radome enclosed paraboloid antenna situated in Finland. The measurements were made with a 1 GHz-band dual beam receiver centred
at 36.8 GHz. The observations are ON-ON measurements, alternating between the source
and the sky in each feed horn. A typical integration time to obtain one flux
density data point is between 1200 and 1400 s. The detection limit at 37 GHz is on the order of 0.2 Jy under optimal conditions. Data
points with a signal-to-noise ratio $<$ 4 are handled as non-detections. The
flux density scale is set by observations of DR 21. Sources NGC 7027, 3C 274
and 3C 84 are used as secondary calibrators. A detailed description of the data reduction and analysis is
given in \citet{terasranta98}. The error on the flux density includes
the contribution from the measurement RMS and the uncertainty of the absolute
calibration. Flux densities at 37 GHz are shown in Fig.~\ref{MWL}.

\subsection{OVRO 40 m}
As part of an ongoing blazar monitoring program, the OVRO 40 m radio telescope has observed PMN J0948$+$0022
at 15~GHz regularly since the end of 2007 \citep{richards11}. This
monitoring program includes about 1700 known and likely $\gamma$-ray loud
blazars above declination $-20^{\circ}$. The sources in this
programme are observed in total intensity twice per week with a 4~mJy
(minimum) and 3\% (typical) uncertainty on the flux densities. Observations were performed
with a dual-beam (each 2.5~arcmin FWHM) Dicke-switched system using
cold sky in the off-source beam as the reference. Additionally, the
source is switched between beams to reduce atmospheric variations. The
absolute flux density scale is calibrated using observations of
3C~286, adopting the flux density (3.44~Jy) from \citet{baars77}. 
This results in about a 5\% absolute scale uncertainty, which is not reflected
in the plotted errors. During the OVRO monitoring, the flux density varied from 671 mJy (on MJD 55725; 2011 June 13) to 335 mJy (on MJD 55836; 2011 October 2), as shown in Fig.~\ref{MWL}.

\subsection{Medicina}

PMN J0948$+$0022 was observed with the 32 m Medicina radio telescope eight times between
2011 May and October. The new enhanced single-dish control acquisition system, which provides enhanced
sensitivity and supports observations with the cross scan
technique, was used. Observations were performed at both 5 and 8.4 GHz; the
typical on-source time was 1.5 minutes and the flux density was calibrated with respect to 3C 286. Since
the signal-to-noise ratio in each scan across the source was low
(typically$\sim 3$), we performed a stacking analysis of the scans, which
allowed us to significantly improve the signal-to-noise ratio and the precision of the
measurement. The flux densities at 5 and 8.4 GHz are reported in Table~\ref{t.medicina}. The peak of the flux density was observed by Medicina
first at 8.4 GHz on 2011 August 1 (MJD 55774), and then at 5 GHz on August 10 (MJD 55783), about 7-8 weeks after the peak observed at
15 GHz by OVRO. Using the Medicina data at 5 GHz and 8.4 GHz together with the
nearest OVRO observation spectral indices of $-$0.54$\pm$0.04,
$-$0.36$\pm$0.03, and $-$0.04$\pm$0.06 were measured for 2011 June 12, August 1, and August 10, respectively,
suggesting a radio spectral evolution in agreement with the behaviour observed
by Effelsberg.

\begin{table}
\begin{center}
\caption{Results of the Medicina radio observations at 5 GHz and 8.4 GHz of PMN J0948$+$0022. \label{t.medicina}}
\begin{tabular}{cccc}
\hline \hline
Date & Date & $S_{5\,\rm GHz}$ & $S_{8.4\,\rm GHz}$ \\
(UT) & (MJD)  & (mJy) & (mJy) \\
\hline
2011-June-12 & 55724 & $0.37 \pm 0.02$ & $0.38 \pm 0.02$ \\
2011-July-29 & 55771 & -- & $0.35 \pm 0.02$ \\
2011-Aug-01  & 55774 & $0.35 \pm 0.02$ & $0.45 \pm 0.02$ \\
2011-Aug-10  & 55783 & $0.45 \pm 0.05$ & $0.39 \pm 0.05$ \\
2011-Sep-08  & 55812 & $0.35 \pm 0.02$ & -- \\
2011-Sep-22  & 55826 & $0.34 \pm 0.02$ & -- \\
2011-Oct-13  & 55847 & $0.33 \pm 0.05$ & -- \\
2011-Nov-16  & 55881 & $0.25 \pm 0.05$ & -- \\
\hline \hline
\end{tabular}
\end{center}
\end{table}

\subsection{MOJAVE: data analysis and results}\label{MOJAVE}

\begin{figure}
\centering
\includegraphics[width=7.5cm]{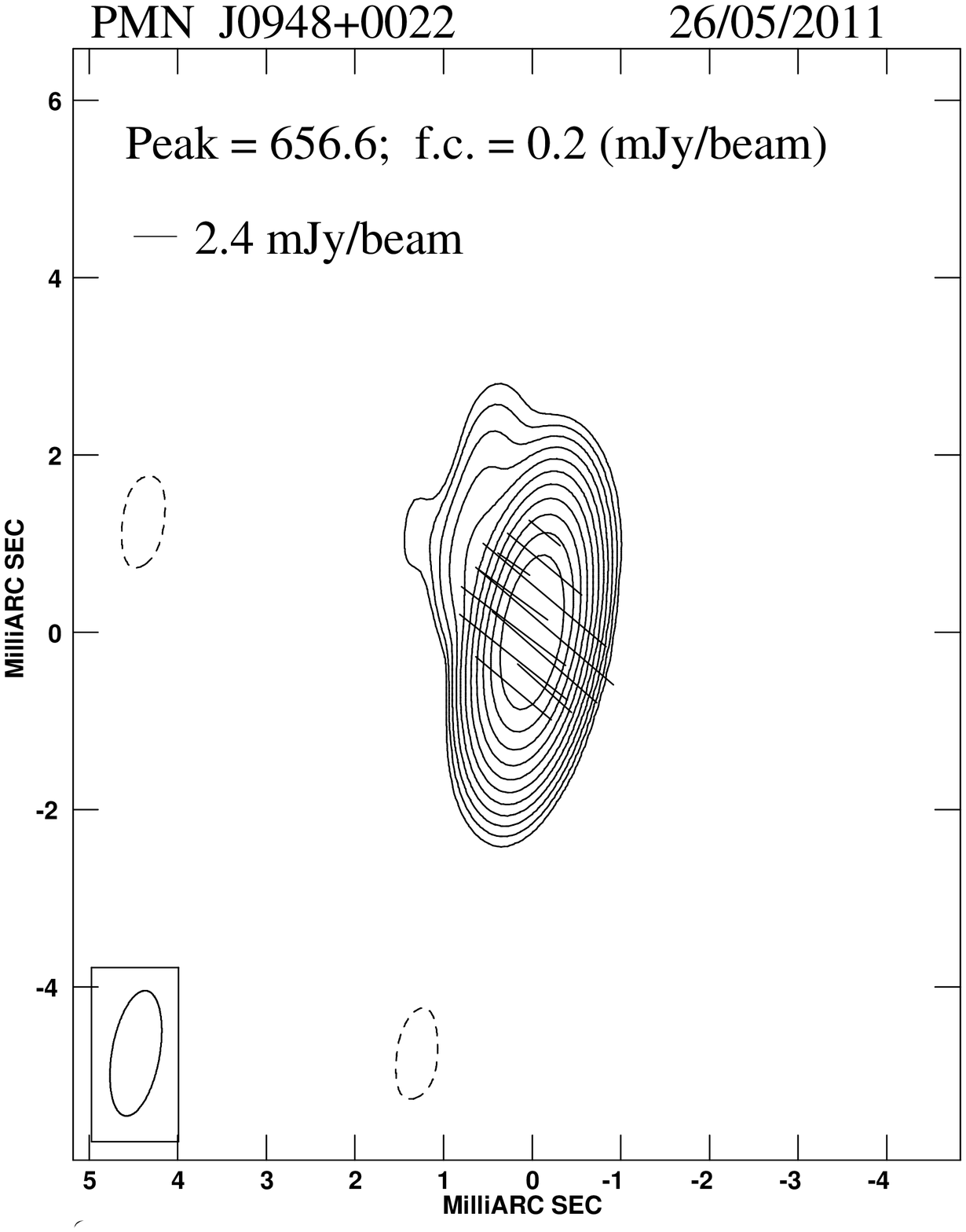}
\caption{VLBA image at 15.3 GHz of PMN J0948$+$0022
collected on 2011 May 26. On the image we provide
the restoring beam, plotted in the bottom-left corner, the
peak flux density in mJy/beam, and the first contour
(f.c.) intensity in mJy/beam, which is three times the
off-source noise level. Contour levels increase by a factor
of 2. The vectors superimposed on the total intensity contours show the percentage
polarization and the position angle of the electric vector.}
\label{VLBA1}
\end{figure}

\begin{figure}
\centering
\includegraphics[width=7.5cm]{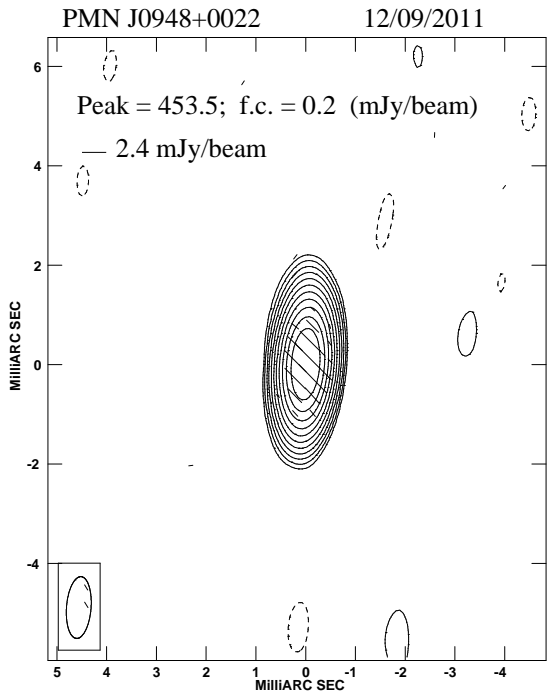}
\caption{VLBA image at 15.3 GHz of PMN J0948$+$0022
collected on 2011 September 12. On the image we provide
the restoring beam, plotted in the bottom-left corner, the
peak flux density in mJy/beam, and the first contour
(f.c.) intensity in mJy/beam, which is three times the
off-source noise level. Contour levels increase by a factor
of 2. The vectors superimposed on the total intensity contours show the percentage
polarization and the position angle of the electric vector.}
\label{VLBA2}
\end{figure}

We investigated the parsec-scale morphology and flux density variability at 15
GHz by means of 5-epoch VLBA data from the {\small MOJAVE} programme
\citep{lister09}. The data sets span the time interval between 2011 February
and December, in order to overlap with the {\em Fermi}-LAT data. We imported
the calibrated {\it uv} data into the National Radio Astronomy Observatory
{\small AIPS} package. In addition
to the total intensity images, we produced the Stokes Q and U images, to
derive information on the polarized emission. The flux density was derived by
means of the {\small AIPS} task JMFIT which performs a Gaussian fit on the
image plane. Total intensity flux density and polarisation information are
reported in Table \ref{tab_moj}. During some observing epochs we detected a
hint of the jet emerging from the core component with a position angle of $\sim$30$^{\circ}$ (see Fig.~\ref{VLBA1} and \ref{VLBA2}), in agreement with previous works
\citep[e.g.,][]{doi06,giroletti11}. In accordance with the flux density observed by OVRO at 15 GHz, a
clear decrease of the flux density was observed by MOJAVE from 657 mJy on
2011 May 26 (MJD 55707) to 378 mJy on 2011 December 12 (MJD 55907). In addition, a higher polarized
emission ($S_{\rm pol}$) and polarisation percentage was observed on May 26 with respect to
December 12 (Table \ref{tab_moj}). On the other hand, the electric vector
position angle (EVPA) of the core does not change significantly, ranging
between 25$^{\circ}$ and 67$^{\circ}$.

\begin{table}
\caption{Flux density and polarisation of PMN\,J0948+0022 from 15 GHz MOJAVE data.}
\begin{center}
\begin{tabular}{cccccc}
\hline\hline
Date & Date &$S_{\rm Core}$&$S_{\rm Jet}$&$S_{\rm pol}$&EVPA\\
(UT) & (MJD) &  (mJy) &  (mJy)   & (mJy) \,\, (\%)& (deg) \\
\hline
2011-02-20 & 55612 & 622 & 3 & 5 \,\, (0.8\%)& 25\\
2011-05-26 & 55707 & 659 & 6 & 13 \,\, (2.0\%)& 51\\
2011-06-24 & 55736 & 665 & 6 & 14 \,\, (2.1\%)& 49\\
2011-09-12 & 55816 & 458 & -- &  6 \,\, (1.3\%)& 44\\
2011-12-12 & 55907 & 380 & -- &  2 \,\, (0.5\%)& 67\\
\hline\hline
\end{tabular}
\label{tab_moj}
\end{center}
\end{table}

\section{Discussion and conclusions}

PMN J0948$+$0022 is the best studied $\gamma$-ray NLSy1 \citep[see e.g.,][]{abdo09c,foschini12}. Simultaneous multiwavelength
observations presented here allow a broad-band characterization of this
source, including the first XMM-{\em Newton} observation of this NLSy1. 

\subsection{Behaviour of the light curves}

In Fig.~\ref{MWL}, we compare the $\gamma$-ray light curve collected by {\em
  Fermi}-LAT in the 0.1--100 GeV energy range to the X-ray (0.3--10 keV), UV
($w2$ filter), optical ($V$, $R$, and $u$ filters), NIR ($J$ and $H$ filters),
and radio (37 GHz and 15 GHz) light curves collected by {\em Swift}, XMM, CRTS,
INAOE, Mets\"ahovi, Effelsberg, and OVRO. Strong variability was observed in $\gamma$ rays,
with two flaring periods peaked on 2011 June 20 and July 28 and a variability amplitude (calculated as the ratio of maximum to minimum flux) of
$\sim$10. Such a variability amplitude as well as the rapid flaring episodes
and high apparent isotropic $\gamma$-ray luminosity ($\sim$10$^{48}$ erg
s$^{-1}$ at the peak) observed in 2011
are typical of FSRQs. During the {\em Swift} observations, PMN~J0948+0022 was observed in an
intermediate X-ray state (0.3--10 keV flux of 3.5--5.9$\times$10$^{-12}$ erg
cm$^{-2}$ s$^{-1}$) between the highest flux observed from this source on 2012 December 30
during a $\gamma$-ray flaring activity \citep[12.6$\times$10$^{-12}$ erg
  cm$^{-2}$ s$^{-1}$; see][]{dammando13} and the low flux observed on 2008 December
8 \citep[2.2$\times$10$^{-12}$ erg cm$^{-2}$ s$^{-1}$;][]{abdo09a}. A clear increase of the flux was observed between 2011 June 14 (MJD 55726) and
July 2 (MJD 55744), after the $\gamma$-ray flare
observed by LAT. The flux increase was accompanied by a hardening of the X-ray
spectrum. However, the lack of X-ray data during the peaks of the $\gamma$-ray activity
does not allow us to draw conclusions regarding a contemporaneous increase of the activity in X-ray and $\gamma$ rays. 

The source increased its $V$-band flux by a factor of $\sim$3 from 2011 May 15
(MJD 55696) to May 26 (MJD 55707). An increse of flux density between 2011 May
24 and 25 of about 50\% in $R$-band was reported by \citet{eggen13}, with a significant
decrease of a factor of $\sim$3.5 in two days. These rapid changes are in agreement
with the optical intraday variability observed in 2011 March 27--31, with a total amplitude
of $\sim$ 0.9 mag during $\sim$4 hours on April 1 \citep{maune13}. Similar intraday
variability has been reported for this source by \citet{liu10, maune11, paliya13} and \citet{itoh13}, indicating
a relativistic jet as the most likely origin for the optical emission in PMN
J0948$+$0022. If this variability is due to the accretion disc, the amplitude
variability should be higher during faint states, when the jet activity is lower, and minimised during bright states, as
already discussed in \citet{maune11}. At the time of the optical peak on 2011
May 25 a high optical polarisation percentage (P=12.31\%) was observed, with a
significant increase with respect to the observation performed in May 24
(P=1.35\%) \citep{eggen13}. An increase of polarisation percentage from 0.8\% on 2011 February 20 to
2.0\% on May 26 was also observed in the MOJAVE data (see Section \ref{MOJAVE}). 

The sparse sampling of the optical and NIR light curves does not allow us to
make a detailed comparison with the $\gamma$-ray and radio light curves. However, it is
worth noting that an increase of flux density was also observed in NIR in
2011 mid-May (about MJD 55700), with no counterpart in $\gamma$ rays. In fact, PMN J0948$+$0022 was quite
active in $H$-band on 2011 April 26 (MJD 55677), and after a decrease in the
following 3 weeks, we observed on 2011 May 22 (MJD 55703) a significant
increase by a factor of $\sim$4.5 in 24 hours. A similar behaviour seems to be observed in $J$-band, with an
increase by a factor of $\sim$2.5 on May 22 (MJD 55703).
On the other hand, no
significant variability was observed in UV during the {\em Swift}-UVOT
observations, probably due to the accretion disc emission that dilutes the
variability in that part of the spectrum.

During the OVRO monitoring the flux density at 15 GHz changed more smoothly
than at higher energies, with an amplitude variability of $\sim$2 between 670 mJy (2011 June 13; MJD 55725) and 335 mJy
(2011 October 2; MJD 55836) and a gradual decrease. A flaring activity
seems to start before the first $\gamma$-ray flare, with a peak at about MJD 55700, close to the optical and NIR flare, and a monotonic
decreasing trend in the following months at 15 and 37 GHz. In the same way the
radio spectra collected by Effelsberg, Medicina, and OVRO showed a
significant spectral evolution with a high frequency component dominating
the emission in 2011 May. A similar spectral evolution was already reported in
\citet{angelakis13} for PMN J0948$+$0022 and other $\gamma$-ray emitting
NLSy1s with variability patterns typical of relativistic jets and thus similar
to blazars. 

The different behaviour observed in the radio-to-optical and the $\gamma$-ray
energy bands could be related to a bending and inhomogeneous jet, as proposed
for some blazars \citep[e.g.,][]{raiteri10,raiteri11}, with a variable misalignment between the zone responsible for the
radio-to-optical emission and the zone responsible for $\gamma$ rays. The change of the
viewing angle of the different emitting regions may produce a change of the
Doppler factor, with the Doppler boosting of the radio-to-optical emission increasing first, followed by an
increase of the Doppler boosting of the $\gamma$-ray emission. This complex
behaviour also could be in agreement with the turbulent extreme multi-cells scenario proposed by \citet{marscher12}.
Alternatively, the radio activity could be
related to the $\gamma$-ray flaring activity observed in 2010 July--August,
but the large delay of about 1 year makes this hypothesis unlikely. The sparse
and irregular sampling, especially from NIR to X-rays, does not allow us to test the different scenarios. 

\subsection{Energetics}

The high apparent isotropic $\gamma$-ray luminosity observed for PMN
J0948$+$0022 in 2011 May--September ($\sim$ 1.8$\times$10$^{47}$ erg s$^{-1}$) should reflect a small viewing angle with respect to the
jet axis and thus high beaming factors, similarly to what is observed for the FSRQs and also for the NLSy1 SBS 0846$+$0513
\citep{dammando12,dammando13b}. This is consistent with the viewing angle of 3$^{\circ}$
used for modeling the SEDs of this source in \citet{foschini11,foschini12}. On
the contrary, most of the radio galaxies detected by the LAT have an apparent
isotropic $\gamma$-ray luminosity lower than 10$^{46}$ erg s$^{-1}$,
suggesting a smaller beaming factor and possibly a different structure of the
jet \citep{abdo10}. Assuming a BH mass of 10$^{8}$ M$_{\odot}$ (but see
Section \ref{mass}) we obtain an Eddington luminosity of 1.3$\times$10$^{46}$ erg s$^{-1}$. During
the 2011 June flare PMN J0948$+$0022 reached an apparent isotropic
$\gamma$-ray luminosity of L$_{\gamma}$ = 8.8$\times$10$^{47}$ erg s$^{-1}$,
making the radiative power L$_{\rm rad}$ = L$_{\gamma}$/$\Gamma^2$ =
3.5$\times$10$^{45}$ erg s$^{-1}$, assuming a quite typical value for this
source of $\Gamma$ = 16 \citep{foschini11,foschini12}. This is about
25\% of the Eddington luminosity, comparable to the values observed for bright FSRQs detected by LAT \citep[see e.g.,][]{nemmen12}.

\subsection{X-ray spectrum}\label{xspectrum}

Thanks to the first high quality XMM-{\em Newton} observation of PMN~J0948$+$0022, we are able to study in detail its
X-ray spectrum (see Section \ref{XMM}). The spectral modelling of the XMM data of PMN J0948$+$0022 showed that emission
from the jet most likely dominates the spectrum above $\sim 2$~keV, while the
emission from the underlying Seyfert galaxy can be seen at lower
energies. Interestingly, the observation of such a component, typical in the
X-ray spectra of radio-quiet NLSy1s, is quite unusual in jet-dominated AGNs, even if not unique \citep[e.g., the FSRQ
  PKS 1510$-$089;][]{kataoka08}. Contrary to what is observed in some blazars
\citep[e.g., BL Lacertae;][]{raiteri10}, no excess absorption above the
Galactic column density was necessary by the fit for modelling the low-energy
part of the spectrum \citep{grupe10}. As well as for PKS 1510$-$089, we hypothesize that the emission below 2
keV was mainly due to the soft X-ray excess. However, we cannot distinguish between different
models for the soft X-ray emission on a statistical basis. Models where the soft emission is partly produced by blurred
reflection, or Comptonisation of the thermal disc emission, or simply a steep
power law, all provide good fits to the data. Our reflection model differs
substantially from that presented by \cite{Bhattacharyya2013}. While we confirm that their model is a good fit to
the spectrum, we note that it relies on a high inclination angle of $i =
74^{\circ}$, which is inconsistent with the blazar-like properties of the
source, as well as the inclusion of a warm absorber, which is not required in any of the other
models. Furthermore, their model does not allow the power law, which illuminates the disc, to also contribute directly to the
spectrum. 

\noindent A blackbody model also gives a comparable fit, but a
temperature of kT = 0.18 keV is necessary. Such a high temperature is
inconsistent with the standard accretion disc theory \citep[see e.g.,][]{shakura73}. Further long-duration X-ray observations with XMM-{\em Newton}, {\em
  Suzaku} and {\em NuSTAR}, preferably when the source is in a low $\gamma$-ray
state, will be needed to place stronger constraints on these models. 

Our interpretation that the X-ray emission above 2 keV is produced by the jet
is also supported by the RMS spectrum (Fig.~\ref{rms}), which shows a break at
1.7~keV, above which the variability increases with energy. While Seyfert
galaxies typically exhibit RMS spectra decreasing with energy above $\sim
1-2$~keV \citep{Markowitz2004,Ponti2007,Chitnis2009}, the opposite behaviour
is often observed in blazars \citep[e.g.,][]{Ravasio2004,Gliozzi2006}. It
should be noted, however, that there is no one-to-one correlation between the
RMS spectrum and AGN type and that RMS spectra often change substantially with
time \citep[as seen in e.g.][]{Gliozzi2006,Larsson2008}. 

The spectrum of PMN J0948$+$0022 changed from a steep slope of $\Gamma$
$\sim$2.1 to a much harder one of $\Gamma$ $\sim$1.5 above 1.7 keV. A similar spectrum has also been observed in the XMM observation of the $\gamma$-ray NLSy1 PKS~2004--447 \citep{gallo06}. An X-ray spectrum unusually
hard for a NLSy1 was observed by {\em Swift}-XRT in SBS 0846$+$513 \citep{dammando12,dammando13b} and PKS 1502$+$036
\citep{dammando13a}, two other NLSy1s detected by LAT. Models more complex than the simple power law were not
applicable in these cases due to insufficient statistics, in particular below
1--2 keV. Thus the 0.3-10 keV spectra collected by {\em Swift}-XRT seem to be dominated by the jet emission.

The small variability amplitude ($\sim$2) observed in X-rays with respect to the $\gamma$ rays
($\sim$10) could be an indication that the X-ray emission is produced by the
low-energy tail of the same electron distribution. On the other hand, the peak flux in X-rays was
observed on 2011 July 2 during a period of low $\gamma$-ray activity,
suggesting that different mechanisms could be at work in the X-ray and
$\gamma$-ray bands (e.g., synchrotron self-Compton and external Compton, respectively). The presence of
the soft X-ray excess below 2 keV could dilute the X-ray variability over the
0.3--10 keV energy range, but an amplitude variability of a factor of
$\sim$2 (with fluxes between 2.1--4.1$\times$10$^{-12}$ erg cm$^{-2}$
s$^{-1}$) was also observed considering only the 2--10 keV energy range.   
In this context no obvious relation exists between the soft X-ray excess and the
$\gamma$-ray emission. An intriguing possibility is that the excess observed
below 2 keV is a signature of the bulk Comptonisation process by a cold
relativistic plasma accelerating along the jet and scattering on disc photons
reprocessed by the broad-line region (BLR) \citep{celotti07}. A dedicated modeling of the source's
SED including XMM-{\em Newton} and {\em Fermi}-LAT data will be presented in a forthcoming paper. 

\subsection{Host galaxy}

The discovery of a relativistic jet in a class of AGN thought to be hosted in spiral
galaxies such as the NLSy1s, as opposed to
blazars and radio galaxies hosted in elliptical galaxies \citep{blandford78},
was a great surprise challenging the current knowledge on how the
jet structures are generated and developed \citep[see e.g.,][]{boett02,marscher10}. Unfortunately only very
sparse observations of the host galaxies of radio-loud NLSy1s are available and the sample of objects studied by \citet{deo06} and \citet{zhou06}
have redshifts $z < 0.03$ and $z < 0.1$, respectively, while four out five of
the NLSy1s detected in $\gamma$ rays have $z > 0.2$. Among the radio-loud
NLSy1s detected up to now by LAT only for the closest one, 1H 0323$+$342, was
the host galaxy clearly detected. Observations with the {\em Hubble Space Telescope} and Nordic Optical Telescope revealed a one-armed galaxy morphology or a
circumnuclear ring, respectively, suggesting two possibilities: the
spiral arm of the host galaxy \cite[]{zhou07} or the residual of a galaxy merger \cite[]{anton08}. These observations, together with the lack of information
about the host galaxy of the other $\gamma$-ray emitting NLSy1s, leaves room for the hypothesis
that the NLSy1s detected in $\gamma$ rays by LAT could have peculiar host galaxies with
respect to the other NLSy1s. Therefore the possibility that the development of relativistic jets in these objects occurs in hosts undergoing strong merger activity, or with non-spiral morphology, cannot be ruled out. Further high-resolution
observations of the host galaxies of PMN J0948$+$0022 and the other $\gamma$-ray
NLSy1s will be fundamental to obtain insights into the onset of production of relativistic jets in these sources.

\subsection{BH mass and jet formation}\label{mass}

The mechanism for producing a relativistic jet is still unclear. In
particular the physical parameters that drive the jet formation are still under
debate. One of the key parameters should be the BH mass, with only large masses allowing an
efficient jet formation \citep[see e.g.,][]{sikora07}. 
Therefore one of the most surprising facts related to the discovery of PMN J0948$+$0022
was the development of a relativistic jet in an object with a relatively small
BH mass, 3.2$\times$10$^{7}$ M$_{\odot}$ \citep{yuan08}. Recently,
\citet{chiaberge11} suggested that a BH mass higher than 10$^{8}$ M$_{\odot}$
is necessary for producing a radio-loud AGN and that the merger history together
with the subsequent galaxy morphology plays a fundamental role. In any case,
the estimated mass of this source, as well as for the other NLSy1s, has large
uncertainties. By means of the broad band SED modeling a BH mass of
1.5$\times$10$^{8}$ M$_{\odot}$ was estimated for PMN J0948$+$0022 in
\citet{foschini11}. \citet{marconi08} suggested that BLR clouds are subjected
to radiation pressure from the absorption of ionizing photons, and by applying a
correction to the virial relation we have higher masses for the NLSy1s. 
Recently, also \citet{calderone13} pointed out that the BH masses of
the NLSy1s estimated by the modelling of optical/UV data with a Shakura \&
Sunyaev disc spectrum could be significantly higher than those derived on the
basis of single epoch virial methods. In particular, for PMN J0948$+$0022 they found a value
of 10$^{9}$ M$_{\odot}$ in agreement with the typical BH mass of blazars. This
may solve the problem of the minimum BH mass predicted in different scenarios of
relativistic jet formation and development, but introduces a possible
new one. For spiral galaxies, the BH mass typically ranges between
  10$^{6}$ and 10$^{8}$ M$_{\odot}$ \citep[see e.g.,][]{woo02}. If the
BH mass is on the larger side of the estimated values, how is it possible to reconcile such a large BH mass with a spiral galaxy?

A second fundamental parameter for the efficiency of relativistic jet production should be the BH
spin, with super-massive black holes (SMBH) in elliptical galaxies having on average much larger spins
than SMBHs in disc-spiral galaxies, as proposed in the ``modified spin paradigm'' \citep{sikora07}. This is because the spin evolution of BHs in spiral galaxies seems to be limited by multiple accretion
events with random orientation of the angular momentum vectors and small increments
of mass, while elliptical galaxies underwent at least one major merger with
large matter accretion triggering an efficient spin-up of the SMBHs. 
Thus, the mass and the spin of the BH seem to be related to the host galaxies, leading to the hypothesis that relativistic jets
can efficiently develop only in elliptical galaxy
\cite[e.g.][]{boett02,marscher10}. However, the presence of a rapidly spinning
BH was inferred by means of X-ray reflection spectroscopy in a few
radio-quiet AGNs hosted by spiral/disc galaxies, suggesting that the BH spin is not the only parameter
that drives the radio-quiet/radio-loud dichotomy \citep[see e.g.,][]{reynolds13}. 

We noted that BH masses of radio-loud NLSy1s reported in \citet{komossa06} and
\citet{yuan08} are generally larger than those in the entire sample of NLSy1s
(M$_{\rm BH}$ $\approx$(2--10)$\times$10$^{7}$ M$_\odot$), even if still
smaller than those in radio-loud quasars. The larger BH masses of radio-loud NLSy1s
could be related to higher mass accretion events that can spin up the BHs. In
the same way the smaller fraction of radio-loud NLSy1s with respect to the radio-loud quasars could
be because the high accretion rate regime does not last sufficiently long in all NLSy1s to substantially spin up the central BH \cite[]{sikora09}. 

Another consideration which is likely to be important for jet formation is the nature of the accretion flow. In particular, a geometrically thick accretion flow is needed in order to create large-scale poloidal magnetic fields, which may play a
dominant role in the launching of jets \citep{reynolds06,sikora13}. In cases where standard thin discs are present, the jet
activity may be due to the dissipation of coronal magnetic fields
\citep{sikora13}. For PMN J0948$+$0022 it is clear that emission from the jet
dominates the X-ray spectrum above 2 keV, while we are likely seeing the
accretion disc plus corona of the AGN at lower X-ray energies. Although we cannot
constrain detailed models for the soft X-ray emission with the current
observations, we note that the spectral slope is similar to that found in radio-quiet NLSy1, indicating that a
standard disc is present as also expected from the high accretion rate. Future
deep X-ray observations with the jet in different states are needed to explore
in detail the connection between the disc and jet in this source. 
\\

The presence of a relativistic jet in some radio-loud NLSy1 galaxies, first
suggested by their variable radio emission and flat spectra, is now confirmed by the {\em Fermi}-LAT detection of five NLSy1s in $\gamma$ rays. PMN
J0948$+$0022 showed all characteristics of the blazar phenomenon with a BH
mass of 10$^{8}$--10$^{9}$ M$_{\odot}$, not much less than those of blazars. The impact on the $\gamma$-ray emission mechanisms of the properties of the central
engine in radio-loud NLSy1s, derived from their peculiar optical
characteristics, is still under debate. In addition, the detection of relativistic jets in a
class of AGN thought to be hosted in spiral galaxies is very intriguing, challenging the theoretical scenario of relativistic jet formation
proposed so far. Further multifrequency observations of this object and other $\gamma$-ray emitting NLSy1s will be fundamental for
investigating in detail their characteristics over the entire electromagnetic spectrum.

\section*{Acknowledgments}

The {\em Fermi} LAT Collaboration acknowledges generous ongoing
support from a number of agencies and institutes that have supported
both the development and the operation of the LAT as well as
scientific data analysis.  These include the National Aeronautics and
Space Administration and the Department of Energy in the United
States, the Commissariat \`a l'Energie Atomique and the Centre
National de la Recherche Scientifique / Institut National de Physique
Nucl\'eaire et de Physique des Particules in France, the Agenzia
Spaziale Italiana and the Istituto Nazionale di Fisica Nucleare in
Italy, the Ministry of Education, Culture, Sports, Science and
Technology (MEXT), High Energy Accelerator Research Organization (KEK)
and Japan Aerospace Exploration Agency (JAXA) in Japan, and the
K.~A.~Wallenberg Foundation, the Swedish Research Council and the
Swedish National Space Board in Sweden. Additional support for science
analysis during the operations phase is gratefully acknowledged from
the Istituto Nazionale di Astrofisica in Italy and the Centre National
d'\'Etudes Spatiales in France.

We thank the {\em Swift} team for making these observations possible, the
duty scientists, and science planners. The OVRO 40 m monitoring program
is supported in part by NASA grants NNX08AW31G and NNX11A043G, and NSF grants AST-0808050 
and AST-1109911. 
This paper is partly based on observations with the 100 m telescope of the
MPIfR (Max-Planck-Institut f\"ur Radioastronomie) at Effelsberg and the Medicina telescope operated by INAF-Istituto di Radioastronomia. We
acknowledge A. Orlati, S. Righini, and the Enhanced Single-dish Control System Development Team. The CSS survey is funded by the National Aeronautics and Space
Administration under Grant No. NNG05GF22G issued through the Science
Mission Directorate Near-Earth Objects Observations Program.  The CRTS
survey is supported by the U.S.~National Science Foundation under
grants AST-0909182. This research has made use of observations obtained with the 2.1 m telescope of the Observatorio Astrofisico Guillermo
Haro (OAGH), in the state of Sonora, Mexico, operated by the Instituto
Nacional de Astrofisica, Optica y Electronica (INAOE), Mexico. OAGH
acknowledges funding from the INAOE Astrophysics Department.
The Mets\"ahovi team acknowledges the support from the Academy of Finland
to our observing projects (numbers 212656, 210338, 121148, and others). This
research made use of data from MOJAVE database that is maintained by the
MOJAVE team (Lister et al. 2009). This work is based on observations obtained
with XMM-{\em Newton}, an ESA science mission with intrument and contributions
directly funded by ESA Member States and the USA (NASA). JL acknowledges financial
support from the Swedish National Space Board. FD thanks A. Breeveld and
P. Roming for useful discussion about OM and UVOT cross-calibration.  JL
thanks Andy Fabian for useful discussion. We thank the anonymous referee, S. Cutini, S. Digel, and D. Thompson for useful comments and suggestions.

\end{document}